\documentclass[prc,twocolumn,superscriptaddress,showpacs,amssymb,amsmath,amsfonts,aps]{revtex4}
\usepackage{graphicx}
\usepackage{dcolumn}
\begin{document}
\title{Measurement of the Polarized Structure Function
$\sigma_{LT^\prime}$ for $p(\vec{e},e'\pi^+)n$ in the $\Delta(1232)$
Resonance Region} 

\newcommand*{\mpaa}{|M_{1+}|^2}
\newcommand*{\mpbb}{|E_{1+}|^2}
\newcommand*{\mpcc}{|S_{1+}|^2}
\newcommand*{\mpdd}{|M_{1-}|^2}
\newcommand*{\mpee}{|E_{0+}|^2}
\newcommand*{\mpff}{|S_{0+}|^2}
\newcommand*{\mpgg}{|S_{1-}|^2}
\newcommand*{\mpa}{M_{1+}}
\newcommand*{\mpb}{E_{1+}}
\newcommand*{\mpc}{S_{1+}}
\newcommand*{\mpd}{M_{1-}}
\newcommand*{\mpe}{E_{0+}}
\newcommand*{\mpf}{S_{0+}}
\newcommand*{\mpg}{S_{1-}}

\newcommand*{\UCONN }{ University of Connecticut, Storrs, Connecticut 06269} 
\affiliation{\UCONN } 

\newcommand*{\VIRGINIA }{ University of Virginia, Charlottesville, Virginia 22901} 
\affiliation{\VIRGINIA } 

\newcommand*{\JLAB }{ Thomas Jefferson National Accelerator Facility, Newport News, Virginia 23606} 
\affiliation{\JLAB } 

\newcommand*{\ASU }{ Arizona State University, Tempe, Arizona 85287-1504} 
\affiliation{\ASU } 

\newcommand*{\SACLAY }{ CEA-Saclay, Service de Physique Nucl\'eaire, F91191 Gif-sur-Yvette, Cedex, France} 
\affiliation{\SACLAY } 

\newcommand*{\UCLA }{ University of California at Los Angeles, Los Angeles, California  90095-1547} 
\affiliation{\UCLA } 

\newcommand*{\CMU }{ Carnegie Mellon University, Pittsburgh, Pennsylvania 15213} 
\affiliation{\CMU } 

\newcommand*{\CUA }{ Catholic University of America, Washington, D.C. 20064} 
\affiliation{\CUA } 

\newcommand*{\CNU }{ Christopher Newport University, Newport News, Virginia 23606} 
\affiliation{\CNU } 

\newcommand*{\DUKE }{ Duke University, Durham, North Carolina 27708-0305} 
\affiliation{\DUKE } 

\newcommand*{\ECOSSEE }{ Edinburgh University, Edinburgh EH9 3JZ, United Kingdom} 
\affiliation{\ECOSSEE } 

\newcommand*{\FIU }{ Florida International University, Miami, Florida 33199} 
\affiliation{\FIU } 

\newcommand*{\FSU }{ Florida State University, Tallahassee, Florida 32306} 
\affiliation{\FSU } 

\newcommand*{\GEISSEN }{ Physikalisches Institut der Universitaet Giessen, 35392 Giessen, Germany} 
\affiliation{\GEISSEN } 

\newcommand*{\GWU }{ The George Washington University, Washington, DC 20052} 
\affiliation{\GWU } 

\newcommand*{\ECOSSEG }{ University of Glasgow, Glasgow G12 8QQ, United Kingdom} 
\affiliation{\ECOSSEG }

\newcommand*{\INFNFR }{ INFN, Laboratori Nazionali di Frascati, Frascati, Italy} 
\affiliation{\INFNFR } 

\newcommand*{\INFNGE }{ INFN, Sezione di Genova, 16146 Genova, Italy} 
\affiliation{\INFNGE } 

\newcommand*{\ISU }{ Idaho State University, Pocatello, Idaho 83209} 
\affiliation{\ISU }

\newcommand*{\ORSAY }{ Institut de Physique Nucleaire ORSAY, Orsay, France} 
\affiliation{\ORSAY } 

\newcommand*{\BONN }{ Institute f\"{u}r Strahlen und Kernphysik, Universit\"{a}t Bonn, Germany} 
\affiliation{\BONN } 

\newcommand*{\ITEP }{ Institute of Theoretical and Experimental Physics, Moscow, 117259, Russia} 
\affiliation{\ITEP } 

\newcommand*{\JMU }{ James Madison University, Harrisonburg, Virginia 22807} 
\affiliation{\JMU } 

\newcommand*{\KYUNGPOOK }{ Kungpook National University, Taegu 702-701, South Korea} 
\affiliation{\KYUNGPOOK } 

\newcommand*{\MIT }{ Massachusetts Institute of Technology, Cambridge, Massachusetts  02139-4307} 
\affiliation{\MIT } 

\newcommand*{\UMASS }{ University of Massachusetts, Amherst, Massachusetts  01003} 
\affiliation{\UMASS } 

\newcommand*{\MOSCOW }{ Moscow State University, General Nuclear Physics Institute, 119899 Moscow, Russia} 
\affiliation{\MOSCOW } 

\newcommand*{\UNH }{ University of New Hampshire, Durham, New Hampshire 03824-3568} 
\affiliation{\UNH } 

\newcommand*{\NSU }{ Norfolk State University, Norfolk, Virginia 23504} 
\affiliation{\NSU } 

\newcommand*{\OHIOU }{ Ohio University, Athens, Ohio  45701} 
\affiliation{\OHIOU } 

\newcommand*{\ODU }{ Old Dominion University, Norfolk, Virginia 23529} 
\affiliation{\ODU } 

\newcommand*{\PENN }{ Penn State University, University Park, Pennsylvania 16802}
\affiliation{\PENN }

\newcommand*{\PITT }{ University of Pittsburgh, Pittsburgh, Pennsylvania 15260} 
\affiliation{\PITT } 

\newcommand*{\ROMA }{ Universita' di ROMA III, 00146 Roma, Italy} 
\affiliation{\ROMA } 

\newcommand*{\RPI }{ Rensselaer Polytechnic Institute, Troy, New York 12180-3590} 
\affiliation{\RPI } 

\newcommand*{\RICE }{ Rice University, Houston, Texas 77005-1892} 
\affiliation{\RICE } 

\newcommand*{\URICH }{ University of Richmond, Richmond, Virginia 23173} 
\affiliation{\URICH } 

\newcommand*{\SCAROLINA }{ University of South Carolina, Columbia, South Carolina 29208} 
\affiliation{\SCAROLINA } 

\newcommand*{\UTEP }{ University of Texas at El Paso, El Paso, Texas 79968} 
\affiliation{\UTEP } 

\newcommand*{\UNIONC }{ Union College, Schenectady, NY 12308} 
\affiliation{\UNIONC } 

\newcommand*{\VT }{ Virginia Polytechnic Institute and State University, Blacksburg, Virginia   24061-0435} 
\affiliation{\VT } 

\newcommand*{\WM }{ College of William and Mary, Williamsburg, Virginia 23187-8795} 
\affiliation{\WM } 

\newcommand*{\YEREVAN }{ Yerevan Physics Institute, 375036 Yerevan, Armenia} 
\affiliation{\YEREVAN }

\newcommand*{\NOWNCATU }{ North Carolina Agricultural and Technical State University, Greensboro, NC 27411}

\newcommand*{\NOWGBGLASGOW }{ University of Glasgow, Glasgow G12 8QQ, United Kingdom}

\newcommand*{\NOWJLAB }{ Thomas Jefferson National Accelerator Facility, Newport News, Virginia 23606}

\newcommand*{\NOWSCAROLINA }{ University of South Carolina, Columbia, South Carolina 29208}

\newcommand*{\NOWFIU }{ Florida International University, Miami, Florida 33199}

\newcommand*{\NOWINFNFR }{ INFN, Laboratori Nazionali di Frascati, Frascati, Italy}

\newcommand*{\NOWOHIOU }{ Ohio University, Athens, Ohio  45701}

\newcommand*{\NOWCMU }{ Carnegie Mellon University, Pittsburgh, Pennsylvania 15213}

\newcommand*{\NOWINDSTRA }{ Systems Planning and Analysis, Alexandria, Virginia 22311}

\newcommand*{\NOWASU }{ Arizona State University, Tempe, Arizona 85287-1504}

\newcommand*{\NOWCISCO }{ Cisco, Washington, DC 20052}

\newcommand*{\NOWdeceased }{ Deceased}

\newcommand*{\NOWUK }{ Kentucky, LEXINGTON, KENTUCKY 40506}

\newcommand*{\NOWSACLAY }{ CEA-Saclay, Service de Physique Nucl\'eaire, F91191 Gif-sur-Yvette, Cedex, France}

\newcommand*{\NOWRPI }{ Rensselaer Polytechnic Institute, Troy, New York 12180-3590}

\newcommand*{\NOWUNCW }{ North Carolina}

\newcommand*{\NOWHAMPTON }{ Hampton University, Hampton, VA 23668}

\newcommand*{\NOWTulane }{ Tulane University, New Orleans, Lousiana  70118}

\newcommand*{\NOWKYUNGPOOK }{ Kyungpook National University, Daegu 702-701, South Korea}

\newcommand*{\NOWCUA }{ Catholic University of America, Washington, D.C. 20064}

\newcommand*{\NOWGEORGETOWN }{ Georgetown University, Washington, DC 20057}

\newcommand*{\NOWJMU }{ James Madison University, Harrisonburg, Virginia 22807}

\newcommand*{\NOWURICH }{ University of Richmond, Richmond, Virginia 23173}

\newcommand*{\NOWCALTECH }{ California Institute of Technology, Pasadena, California 91125}

\newcommand*{\NOWMOSCOW }{ Moscow State University, General Nuclear Physics Institute, 119899 Moscow, Russia}

\newcommand*{\NOWVIRGINIA }{ University of Virginia, Charlottesville, Virginia 22901}

\newcommand*{\NOWYEREVAN }{ Yerevan Physics Institute, 375036 Yerevan, Armenia}

\newcommand*{\NOWRICE }{ Rice University, Houston, Texas 77005-1892}

\newcommand*{\NOWINFNGE }{ INFN, Sezione di Genova, 16146 Genova, Italy}

\newcommand*{\NOWBATES }{ MIT-Bates Linear Accelerator Center, Middleton, MA 01949}

\newcommand*{\NOWODU }{ Old Dominion University, Norfolk, Virginia 23529}

\newcommand*{\NOWVSU }{ Virginia State University, Petersburg,Virginia 23806}

\newcommand*{\NOWORST }{ Oregon State University, Corvallis, Oregon 97331-6507}

\newcommand*{\NOWMIT }{ Massachusetts Institute of Technology, Cambridge, Massachusetts  02139-4307}

\newcommand*{\NOWCNU }{ Christopher Newport University, Newport News, Virginia 23606}

\newcommand*{\NOWGWU }{ The George Washington University, Washington,
DC 20052}

\newcommand*{\NOWSAK } {Sakarya University, Sakarya, Turkey} 

  
\author{K.~Joo}
     \affiliation{\UCONN}
\author{L.C.~Smith}
     \affiliation{\VIRGINIA}
\author{I.G.~Aznauryan}
     \affiliation{\YEREVAN}
\author{V.D.~Burkert}
     \affiliation{\JLAB}
\author{R.~Minehart}
     \affiliation{\VIRGINIA}
\author{G.~Adams}
     \affiliation{\RPI}
\author{P.~Ambrozewicz}
     \affiliation{\FIU}
\author{E.~Anciant}
     \affiliation{\SACLAY}
\author{M.~Anghinolfi}
     \affiliation{\INFNGE}
\author{B.~Asavapibhop}
     \affiliation{\UMASS}
% ...
\collaboration{The CLAS Collaboration}

\noaffiliation
\begin{abstract}
{The polarized longitudinal-transverse structure function $\sigma_{LT^\prime}$ 
has been measured using the $p(\vec e,e'\pi^+)n$ reaction in the $\Delta(1232)$ 
resonance region at $Q^2=0.40$ and $0.65$~GeV$^2$. No previous $\sigma_{LT^\prime}$ 
data exist for this reaction channel.  The kinematically complete experiment was 
performed at Jefferson Lab with the CEBAF Large Acceptance Spectrometer (CLAS) 
using longitudinally polarized electrons at an energy of 1.515 GeV. 
A partial wave analysis of the data shows generally better agreement with 
recent phenomenological models of pion 
electroproduction compared to the previously measured $\pi^0 p$ channel.  
A fit to both $\pi^0 p$ and $\pi^+ n$ channels using a unitary isobar model 
suggests the unitarized Born terms provide a consistent description of
the non-resonant background. The $t$-channel pion pole term is  
important in the $\pi^0 p$ channel through a rescattering correction,  
which could be model-dependent.} 
\end{abstract}
\maketitle

The excitation of nucleon resonances using electromagnetic interactions
is an essential tool for understanding quark confinement. However, the 
excited states of the nucleon decay rapidly through emission of mesons.  Thus, the 
resonance formation mechanism can involve both hadronic structure and reaction dynamics,
intermixing quark and meson degrees of freedom.  To understand the role of
the meson cloud in resonance photoexcitation, a variety of theoretical approaches 
have been developed, e.g. chiral-quark and soliton models, chiral perturbation theory, 
dispersion relations, effective Lagrangian
and dynamical models, and most recently, lattice QCD. 
 
\begin{figure}[h]
\includegraphics[scale=0.45]{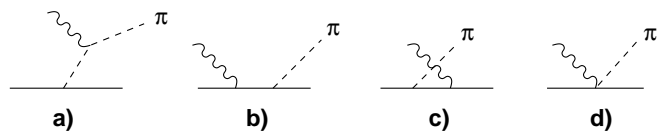}
\caption{Born terms which contribute to
non-resonant background in $\pi$ electroproduction: a) $t-$channel pion pole, b) $s-$channel nucleon pole, c) $u-$channel nucleon pole, and d) contact term.}
\label{fig:fig1}
\end{figure}

A new generation of high-precision photo- and electroproduction 
experiments have made it 
possible to test theoretical predictions with unprecedented accuracy.
The most precise measurements exist for excitation energies around the $\Delta(1232)$ 
resonance and four-momentum transfers $Q^2 < 1 $~GeV$^2$.  Experiments using polarized 
real photons at LEGS and Mainz \cite{bla97,bec97} 
and unpolarized electrons at Bates, ELSA and Jefferson Lab \cite{fro99,mer01,joo02,spar03} 
have measured $\Delta^+ \rightarrow p\pi^0$ 
decay angular distributions with the goal of determining the magnitude
and $Q^2$ evolution of the $N\Delta$ transition photocoupling amplitudes.  However, theoretical calculations 
predict a substantial modification of the $N\Delta$ form factors due to the presence 
of non-resonant Born diagrams (Fig.~\ref{fig:fig1}). 
Moreover, these predictions are subject to considerable model dependence from the 
treatment of $\pi N$ rescattering in the final state. 

To better study these non-resonant contributions, several recent $p(\vec e,e'p)\pi^0$ experiments 
in the $\Delta(1232)$ region \cite{war98,pos01,bar02,kunz03,bis03,joo03} 
have utilized single-spin polarization observables to directly determine the imaginary 
part of interfering amplitudes.  In this way, the 
non-resonant amplitudes, which are largely real, are greatly amplified by the 
imaginary part of the dominant $\Delta(1232)$ 
$M_{1+}^{3/2}$ resonant multipole. Until now, beam asymmetry measurements existed only
for the $\pi^0 p$ channel, where pion rescattering corrections are large and 
model-dependent \cite{geh70,craw71}.
Predictions for the $\pi^+ n$ channel show less model dependence, and are dominated by 
the $\it{t}$-channel pion pole and contact Born terms, which are absent or weak in the
$\pi^0 p$ channel.  Measurement of both charge channels is therefore essential to
test the consistency of the model descriptions.
 
We present the first measurements of the longitudinal-transverse
polarized structure function  $\sigma_{LT^\prime}$ obtained 
in the $\Delta(1232)$ resonance region using the $p(\vec e,e'\pi^+)n$ reaction.  
The data reported here span the invariant mass interval $W= 1.1-1.3$ GeV at 
$Q^2=0.40$ and $0.65$ GeV$^2$, and cover the 
full angular range in the $\pi^+ n$ center-of-mass (c.m.). These data were taken simultaneously with
the $p(\vec e,e'p)\pi^0$ channel for which results were reported previously \cite{joo03}. 

The experiment was performed at the Thomas Jefferson National Accelerator Facility (Jefferson Lab)
using a 1.515 GeV, 100\% duty-cycle beam of longitudinally 
polarized electrons incident on a liquid-hydrogen target. 
The electron polarization was determined by M{\o}ller polarimeter measurements to be 
$0.690\pm0.009$(stat.)$\pm0.013$(syst.). Scattered electrons and pions were detected in
the CLAS spectrometer~\cite{mec03}. Electron triggers were enabled
through a hardware coincidence of the gas C\v{e}renkov
counters and the lead-scintillator electromagnetic calorimeters.  Particle identification
was accomplished using momentum reconstruction in the tracking system
and time of flight from the target to the scintillators. Software 
fiducial cuts were used to exclude regions of non-uniform detector response.
Kinematic corrections were applied to compensate for drift chamber
misalignments and uncertainties in the magnetic field.  The $\pi^+ n$ final state was identified using a $2\sigma$ cut on 
the missing neutron mass. Target window backgrounds were suppressed with cuts 
on the reconstructed $e'\pi^+$ target vertex. 
 
\begin{figure}[h]
\includegraphics[scale=0.42]{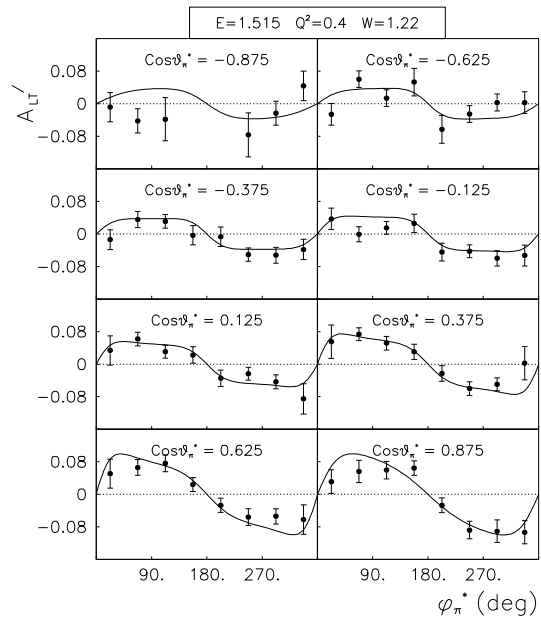}
\caption{CLAS measurement of the beam asymmetry $A_{LT^\prime}$ versus $\phi^*_\pi$ 
for the $p(\vec e,e'\pi^+)n$ reaction at
$Q^2$=0.40~GeV$^2$ and $W=1.22$~GeV. Bin sizes were $\Delta Q^2=0.2~$GeV$^2$ and $\Delta W=0.04~$GeV.  The curves show predictions
from the MAID2000 model described in the text.}
\label{fig:fig2}
\end{figure}

The single pion electroproduction cross section is given by:
\begin{equation}
\frac{d\,^4\sigma^h}{dQ^2 dW d\Omega^*_{\pi}} = J\,\Gamma_v\,\frac{d\,^2\sigma^h}{d\Omega^*_{\pi}},
\end{equation}
where $\Gamma_v$ is the virtual photon flux and the 
Jacobian $J = \partial(Q^2,W)/\partial(E',\cos\,\theta_e,\phi_e)$ relates the differential
volume element $dQ^2 dW$ of the binned data to the measured electron kinematics 
$dE'\,d\cos\,\theta_e\,d\phi_e$.  Here $d\,^2\sigma^h$ is the c.m. differential cross
section for $\gamma^* p \rightarrow n \pi^+$ with electron beam helicity ($h=\pm1$). For an unpolarized target $d\,^2\sigma^h$ depends 
on the transverse $\epsilon$ and longitudinal $\epsilon_L$  polarization of the virtual 
photon through five structure functions: $\sigma_T,\sigma_L,\sigma_{TT}$ and the 
transverse-longitudinal interference terms $\sigma_{LT}$ and $\sigma_{LT^\prime}$:
\begin{eqnarray}
\frac{d\,^2\sigma^h}{d\Omega^*_{\pi}} &=& \frac{p^*_{\pi}}{k_{\gamma}^*} (\sigma_{0} +
h\sqrt{2\epsilon_L(1-\epsilon)}\,\sigma_{LT^\prime}\,\sin\,\theta^*_{\pi}\,\sin\,\phi^*_{\pi}),  \nonumber 
\\
\sigma_{0} &=& \sigma_T+\epsilon_L\sigma_L+\epsilon\,\sigma_{TT}\,\sin^2\theta^*_{\pi}\,\cos\,2\phi^*_{\pi} \nonumber \\
~&+&
\sqrt{2\epsilon_L(1+\epsilon)}\,\sigma_{LT}\,\sin\,\theta^*_{\pi}\,\cos\,\phi^*_{\pi},
\label{eq:str}
\end{eqnarray}
where $p_{\pi}^*$ and $\theta^*_{\pi}$ are the $\pi^+$ c.m. momentum and polar angle, 
$\phi^*_{\pi}$ is the azimuthal rotation of the hadronic plane with respect to the electron
scattering plane, $\epsilon = (1+2|\vec{q}\,|^2\,\tan^2(\theta_e/2)/Q^2)^{-1}$, $\epsilon_L=(Q^2/|k^*|^2)\epsilon$, $|k^*|$ is the virtual photon c.m. momentum, and $k_{\gamma}^*$ is
the real photon equivalent energy.

Determination of $\sigma_{LT^\prime}$ was made through the asymmetry $A_{LT^\prime}$: 
\begin{eqnarray}
A_{LT^\prime} &=&\frac{d\,^2\sigma^+ - d\,^2\sigma^-}{d\,^2\sigma^+ +
d\,^2\sigma^-}  \\ 
&=&
\frac{\sqrt{2\epsilon_L(1-\epsilon)}\,\sigma_{LT^\prime}\,\sin\,\theta^*_{\pi}\,\sin\,\phi^*_{\pi}}{\sigma_{0}}.
\label{eq:altp}
\end{eqnarray}
The asymmetry $A_{LT^\prime}$ 
was obtained for individual bins of $(Q^2,W,\cos\theta_{\pi}^*,\phi_{\pi}^*)$
by dividing the measured single-spin beam asymmetry $A_m$ by the
magnitude of the electron beam polarization $P_e$:
\begin{eqnarray}
A_{LT^\prime} &=& \frac{A_m}{P_e}  \\
A_m &=& \frac{N_\pi\,^+ - N_\pi\,^-}
{N_\pi\,^+ + N_\pi\,^-}, 
\label{eq:altp_m}
\end{eqnarray}
where $N_{\pi}^{\pm}$ is the number of detected $n \pi^+$ events for each
electron beam helicity state, normalized to beam charge. 
Acceptance studies which varied the sizes of all kinematic bins 
showed no significant helicity dependence, leaving $A_m$ largely 
free from systematic errors. Radiative corrections were 
applied for each bin using the program 
recently developed by Afanasev {\it et al.} for exclusive pion 
electroproduction~\cite{aku02}. Corrections were 
also applied to compensate for cross section variations over the width
of each bin, using the cross section model MAID2000, described below.
An example of the measured $\phi^*_\pi$ dependence of $A_{LT^\prime}$ 
is shown in Fig.~\ref{fig:fig2}.  Next the $A_{LT^\prime}$ distributions
were multiplied by the unpolarized $p(e,e'\pi^+)n$ cross section $\sigma_{0}$,
using a parameterization of measurements of $\sigma_{0}$ made during the
same experiment~\cite{hov01}.  The structure function 
$\sigma_{LT^\prime}$ was then extracted using Eq.~\ref{eq:altp} by fitting 
the $\phi_{\pi}^*$ distributions. 
Systematic errors for $\sigma_{LT^\prime}$ were dominated by 
uncertainties in determination of the electron beam polarization 
and the parameterization of $\sigma_{0}$. The systematic error for $A_m$ is 
negligible in comparison. Quadratic addition of the individual
contributions yields a total relative systematic error of $< 6\%$ for all of
our measured data points.

\begin{figure}[h]
\includegraphics[scale=0.47]{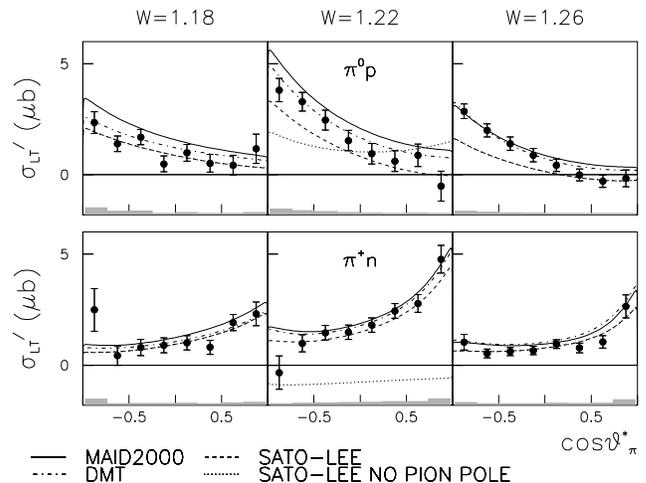}
\caption{CLAS measurements of $\sigma_{LT^\prime}$ versus $\cos\theta^*_\pi$ 
for the $\pi^0 p$ channel~\cite{joo03}~(top) and for the $\pi^+ n$ channel (bottom) extracted at
$Q^2$=0.40~GeV$^2$ and $W=1.18-1.26$~GeV. The curves show model predictions discussed
in the text.  The shaded bars show estimated systematic errors.}
\label{fig:fig3}
\end{figure}

Figure~\ref{fig:fig3} shows typical c.m. angular distributions for  $\sigma_{LT^\prime}$ 
at $Q^2$=0.40~GeV$^2$ and $W=1.18-1.26$~GeV. Our previous measurement  
for the $\pi^0 p$ \cite{joo03} channel (top) and our new measurement for 
the $\pi^+ n$ channel (bottom) are shown compared to phenomenological models by
Sato and Lee (SL)~\cite{sat01}, the Dubna-Mainz-Taipei
(DMT) group~\cite{kam99}, and Drechsel (MAID)\cite{dre99}. These models combine 
Breit-Wigner type resonant amplitudes with backgrounds arising from 
Born diagrams and $\it{t}$-channel vector-meson exchange, while different methods 
are used to satisfy unitarity.  The SL and DMT models use a reaction theory 
to calculate the effect of off-shell $\pi N$ rescattering.  
MAID uses a $K$-matrix approximation, by incorporating the $\pi N$ scattering 
phase shifts \cite{arnd04} into the background amplitudes and treating the rescattered
pion as on-shell.  All well-established resonances are included in DMT and MAID2000, whereas 
SL treats only the $\Delta(1232)$.

The measured angular distributions of $\sigma_{LT^\prime}$
for the $\pi^+ n$ channel show a 
strong forward peaking for $W$ bins around the $\Delta(1232)$, 
in contrast to the $\pi^0 p$ channel, which shows backward peaking.
The calculations qualitatively describe the peaking behavior of both
the $\pi^0$ and $\pi^+$ channels, which
arises largely from the pion pole term (Fig.~\ref{fig:fig1} and SL curves
on Fig.~\ref{fig:fig3}), as 
discussed shortly.  The largest variation between the models occurs in
their predictions for the overall magnitude of $\sigma_{LT^\prime}$, 
although the variation is substantially smaller for the $\pi^+ n$ channel. 

A more quantitative comparison was made through fitting the
extracted $\sigma_{LT^\prime}$ angular distributions using the Legendre 
expansion:
\begin{eqnarray}
\sigma_{LT^\prime} & = & D_0^\prime + D_1^\prime P_1(\cos\theta^*_\pi)
+  D_2^\prime P_2(\cos\theta^*_\pi),
\label{eq:str_legendre} 
\end{eqnarray} 
where $P_l(\cos\theta^*_\pi)$ is the $l^{th}$-order Legendre polynomial
and $D_l^\prime$ is the corresponding Legendre moment. 
For single pion electroproduction, each moment can be written as
an expansion in magnetic $(M_{l_\pi\pm})$, 
electric $(E_{l_\pi\pm})$, and scalar $(S_{l_\pi\pm})$ $\pi N$ multipoles~\cite{ras89}: 
\begin{eqnarray}
D_{0}^\prime &=& -{Im}((M_{1-} - M_{1+} +3E_{1+})^*S_{0+} \nonumber \\
 &+& E_{0+}^*(S_{1-}-2S_{1+}) + ...) \\
D_{1}^\prime &=& -6{Im}((M_{1-} - M_{1+} +3E_{1+})^*S_{1+} \nonumber \\
 &+& E_{1+}^*(S_{1-}-2S_{1+}) + ...) \\
D_2^\prime & = & -12{Im}((M_{2-} -  E_{2-})^*S_{1+} \nonumber \\
 &+& 2E_{1^+}^*S_{2^-} + ...),
\end{eqnarray}
where the $\pi N$ angular momentum $l_{\pi}$ combines with the nucleon spin to
give the total angular momentum $J=l_\pi \pm \frac{1}{2}$.
The expansion is truncated at $l_\pi$=2,
since interference terms involving the resonant multipoles $M_{1+}$, $E_{1+}$ and $S_{1+}$
dominate near the $\Delta(1232)$. 

\begin{figure}[h]
\includegraphics[scale=0.45]{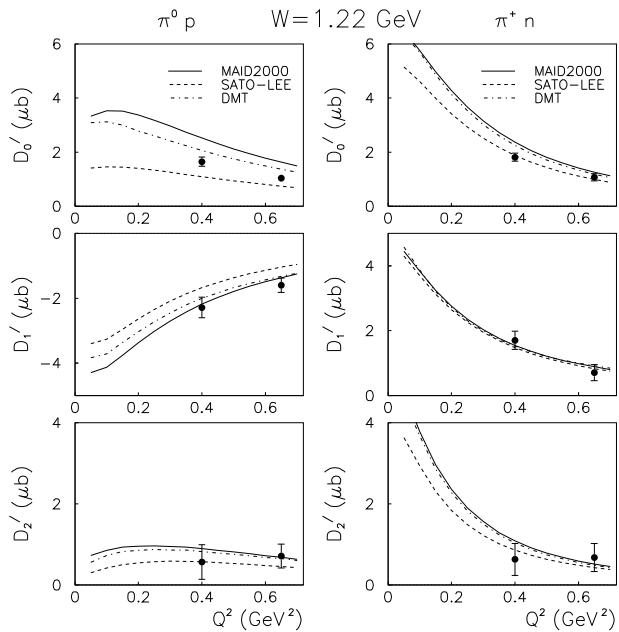}
\caption{The $Q^2$-dependence of Legendre moments of $\sigma_{LT^\prime}$ 
for the $\pi^0 p$ channel~\cite{joo03}~(left) and $\pi^+ n$ channel (right).  
The curves show model predictions described in the text.  
The data points are CLAS measurements showing
statistical errors only.}
\label{fig:fig4}
\end{figure}

Fig.~\ref{fig:fig4} shows the model predictions for the $Q^2$ dependence of the Legendre 
moments at $W=1.22$~GeV compared to our measurements at
$Q^2$=0.4 and 0.65~GeV$^2$. In contrast to our previous result for 
$D_0^\prime$($\pi^0 p$)~\cite{joo03}, which strongly disagreed with the MAID2000 and SL 
predictions, our result for $D_0^\prime(\pi^+ n)$ is much closer to those models.
The model variation is less pronounced, although the SL curve is still lower than the rest, due
to the much smaller $S_{0+}$ multipole in this model.  Good agreement occurs for 
$D_1^\prime$($\pi^+ n$), where there is almost no model dependence in the predictions.  
In contrast, $D_1^\prime$($\pi^0 p$) shows more model dependence, with our measurement 
favoring MAID2000.  For $D_2^\prime$, our results are consistent with the model 
predictions in sign and overall magnitude, although with large statistical errors.
 
\begin{figure}[h]
\includegraphics[scale=0.45]{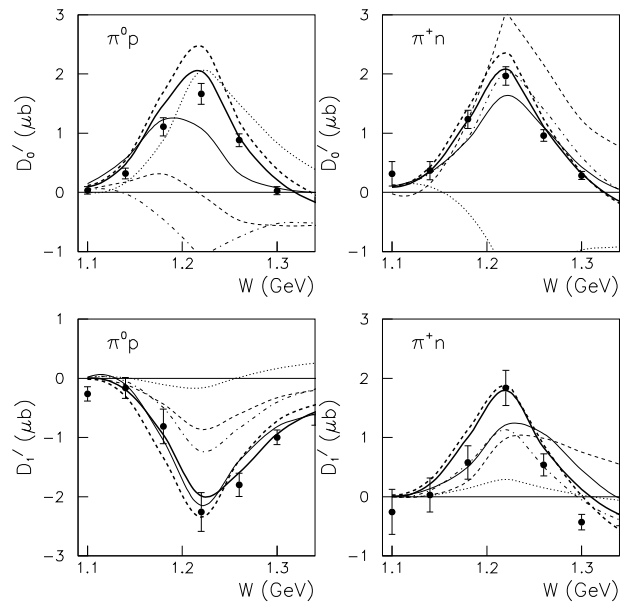}
\caption{The $W$-dependence of $\sigma_{LT^\prime}$ Legendre moments 
$D_0^\prime$ and $D_1^\prime$ 
for $\pi^0 p$~\cite{joo03}~(left) and $\pi^+ n$ (right) at $Q^2=0.4$~GeV$^2$.  The bold 
curves show isobar model fits: UIM (solid), MAID2003 (dashed).  The solid thin curve
shows the SAID XF18/SM01 fit.  The other curves show 
the UIM with the electric (dashed), magnetic (dot-dashed),
and pion pole (dotted) Born terms turned off.  }
\label{fig:fig5}
\end{figure}

The published electroproduction database is undergoing analysis by several groups in 
order to better determine the $Q^2$ dependence of 
the resonant multipoles which contribute to Eqs.~8-10.  The
MAID2003 fit \cite{tiat03} includes recent $\pi^0$ electroproduction data from
Mainz, Bates, Bonn and JLAB, while the more comprehensive SAID analysis \cite{said01} 
includes all previously published $\pi^0$ and $\pi^+$ data.  Finally the unitary 
isobar model (UIM) of Aznauryan~\cite{azn03} 
was fitted solely to the CLAS $\pi^0$ and $\pi^+$ electroproduction data (including the current
polarization data) at $Q^2=0.4$ and 0.65~GeV$^2$.  Fig.~\ref{fig:fig5} 
shows these fits compared to the $W$ dependence of the measured 
Legendre moments, $D_0^\prime$ and $D_1^\prime$. 
  
The UIM fits show the best overall agreement with the $\sigma_{LT^\prime}$ data, especially
in the $\pi^+n$ channel, while  
MAID2003 still overpredicts $D_0^\prime(\pi^0p)$.  This may be due 
to the lack of polarization data in the global MAID fit. However the 
UIM fit also overshoots $D_0^\prime(\pi^0p)$ slightly below the $\Delta(1232)$. 
The SAID XF18/SM01 solution \cite{igor01} shows a somewhat different $W$ dependence compared
to the isobar models, which may reflect the different method of
unitarization used in the SAID approach.  

To explore the sensitivity of this polarization observable to backgrounds,
we turned off various Born terms in the UIM calculation.  
For example, in Fig.~\ref{fig:fig5} it is seen that the $\pi^+ n$ channel is 
strongly sensitive to the $\it{t}$-channel pion pole term,  while 
$D_0^\prime(\pi^0p)$ is similarly affected by the $\it{s}-$ and $\it{u}-$channel 
electric and magnetic Born diagrams.  Therefore, small adjustments to the 
hadronic form factors or meson couplings for these diagrams can affect
the fits.  The $\it{t}$-channel pion pole diagram is surprisingly 
important for $D_1^\prime(\pi^0p)$, where it strongly affects the phases of
the $S_{1+}$ and $E_{1+}$
multipoles \cite{geh70} which are responsible for much of the predicted backward peaking in Fig.~\ref{fig:fig3}.  
This was also verified by turning off the pole term in the SL model
(dotted curve in Fig.~\ref{fig:fig3}).
Note the pion pole can only influence the $\pi^0 p$ channel as a rescattering
correction \cite{craw71} via $\pi^+ n \rightarrow \pi^0 p$, which is introduced using the 
$K$-matrix method in UIM and MAID, or through an explicit meson-exchange
potential in dynamical models.

The generally good agreement of the UIM fits to both our $\pi^+$ and $\pi^0$ data suggests
that the $K$-matrix method of unitarizing the Born terms provides a
consistent description of the backgrounds in the $\Delta(1232)$ region.  
More polarization data is needed at lower $Q^2$, 
which will allow further study 
of the $D_0^\prime(\pi^0p)$ term in a region 
where model sensitivity to pion rescattering is greatest. 

We acknowledge the efforts of the staff of the Accelerator and Physics Divisions at 
Jefferson Lab in their support of this experiment.  This work was supported in
part by the U.S. Department 
of Energy and National Science Foundation, the Emmy Noether Grant from the 
Deutsche Forschungsgemeinschaft, the French Commissariat a l'Energie 
Atomique, the Italian Istituto Nazionale di Fisica Nucleare, and the Korea Research 
Foundation. The Southeastern Universities Research Association (SURA) 
operates the Thomas Jefferson Accelerator Facility for the United States Department 
of Energy under contract DE-AC05-84ER40150.


\end{document}